\documentclass{bvm} 
\usepackage{hyperref} 
\usepackage{comment} 
\begin{document}
\newcommand{\bvmyear}{2026}
\selectlanguage{english} 
\title{Bridging Radiology and Pathology: A DICOM-Based Framework for Multimodal Mapping and Integrated Visualization}

\titlerunning{Unified Viewer for Radiology and Pathology}

\newcounter{equalSeniorCounter}
\newcommand\equalSenior{
	\renewcommand{\thefootnote}{$\ddagger$}
	\ifnum \value{equalSeniorCounter}=0
		\footnote{These authors contributed equally to this work as co-senior authors.}%
		\setcounter{equalSeniorCounter}{1}%
	\else
		\footnotemark[2]
	\fi
	\renewcommand{\thefootnote}{\arabic{footnote}}%
	\setcounter{footnote}{0}
}
\author{
	\fname{Nilesh P.} \lname[0009-0007-9376-1456]{Rijhwani} \inst{1,2} \affiliation{German Cancer Research Center} \authorsEmail{nileshparshotam.rijhwani@dkfz-heidelberg.de} \street{Im Neuenheimer Feld} \housenumber{223} \zipcode{69120} \city{Heidelberg} \country{Germany} \isResponsibleAuthor,
	\fname{Titus J.} \lname[0000-0002-3620-5919]{Brinker} \inst{1} \affiliation{German Cancer Research Center} \authorsEmail{titus.brinker@dkfz.de} \street{Im Neuenheimer Feld} \housenumber{223} \zipcode{69120} \city{Heidelberg} \country{Germany},
    \fname{Peter} \lname[0000-0002-5285-7554]{Neher} \inst{2} \affiliation{German Cancer Research Center} \authorsEmail{p.neher@dkfz-heidelberg.de} \street{Im Neuenheimer Feld} \housenumber{223} \zipcode{69120} \city{Heidelberg} \country{Germany},
    \fname{Marco} \lname[0000-0001-9629-0564]{Nolden} \inst{2} \affiliation{German Cancer Research Center} \authorsEmail{m.nolden@dkfz-heidelberg.de} \street{Im Neuenheimer Feld} \housenumber{223} \zipcode{69120} \city{Heidelberg} \country{Germany},
	\fname{Klaus} \lname[0000-0002-6626-2463]{Maier-Hein} \inst{2,3,4,5,6} \affiliation{German Cancer Research Center} \authorsEmail{k.maier-hein@dkfz-heidelberg.de} \street{Im Neuenheimer Feld} \housenumber{223} \zipcode{69120} \city{Heidelberg} \country{Germany},
	\fname{Maximilian} \lname[0009-0001-8305-6370]{Fischer} \inst{2,3,4,5}\affiliation{German Cancer Research Center} \authorsEmail{maximilian.fischer@dkfz-heidelberg.de} \street{Im Neuenheimer Feld} \housenumber{223} \zipcode{69120} \city{Heidelberg} \country{Germany}\equalSenior,
	\fname{Christoph} \lname[0000-0001-7136-298X]{Wies}\inst{1,3}\affiliation{German Cancer Research Center} \authorsEmail{christoph.wies@dkfz.de}\equalSenior \street{Im Neuenheimer Feld} \housenumber{223} \zipcode{69120} \city{Heidelberg} \country{Germany}
}

\authorrunning{Rijhwani et al.}

\institute{
\inst{1} Division of Digital Prevention, Diagnostics and Therapy Guidance, German Cancer Research Center (DKFZ), Heidelberg, Germany\\
\inst{2} Division of Medical Image Computing , German Cancer Research Center (DKFZ), Heidelberg, Germany\\
\inst{3} Medical Faculty, University Heidelberg, Heidelberg, Germany\\
\inst{4} Research Campus M\textsuperscript{2}OLIE, Mannheim, Germany \\ 
\inst{5} German Cancer Consortium, DKFZ Core Center Heidelberg, Heidelberg, Germany\\
\inst{6} Pattern Analysis and Learning Group, Department of Radiation Oncology, Heidelberg University Hospital, Heidelberg, Germany\\
\inst{6} Pattern Analysis and Learning Group, Department of Radiation Oncology, Heidelberg University Hospital, Heidelberg, Germany\\
}

\email{maximilian.fischer@dkfz-heidelberg.de}

\maketitle

\begin{abstract}
Accurate disease diagnosis depends on effective collaboration between medical specialties, yet departments often use distinct data systems and proprietary formats. This heterogeneity hinders joint analysis and integration of complementary diagnostic information. The use of separate viewers for each modality further restricts cross-specialty collaboration. Although multimodal integration, particularly between radiology and pathology, has demonstrated potential for identifying novel biomarkers, it still relies heavily on manual, time-consuming data pairing. This project introduces an interdisciplinary toolbox that can operate within the Kaapana framework or as a standalone tool to bridge radiology and pathology. By linking modality-specific viewers and extending them with automated image registration and alignment, the platform enables efficient, scalable multimodal analysis. The integrated environment promotes reproducible workflows, accelerates cross-disciplinary research, and facilitates deeper insights into disease mechanisms and patient care.
\end{abstract}

\section{Introduction}
The growing availability of multimodal medical data creates new opportunities for integrated diagnostic and research workflows, but also introduces significant challenges in data management, spatial alignment, and visualization. In particular, the combined assessment of radiological and pathological data holds high potential for comprehensive disease characterization. However, such integration remains technically demanding due to fundamental differences in scale, modality, and acquisition context. Although mature visualization frameworks exist for radiology and, increasingly, for digital pathology, few systems support their spatially registered and interactive inspection within a unified environment. Radiology has undergone extensive digitization, supported by standardized Digital Imaging and Communications in Medicine (DICOM) based infrastructures and well-established analytical tools \cite{ohif2020}. In contrast, digital pathology is still transitioning toward routine digital adoption, and specialized microscopy viewers often operate in isolation from radiological systems \cite{slim_idc}. This fragmentation limits consistent multimodal exploration and constrains the interpretability of algorithmic pipelines that depend on joint visualization and spatial correspondence between imaging domains.

In this paper, we present an approach that builds on the complementary principle of extending existing ecosystems rather than constructing new infrastructure. We developed a lightweight, modular viewer that enables the synchronized visualization of spatially registered radiological and histopathological data. The system integrates open-source components for data management, registration, and rendering into a single, coherent web interface. It supports standard formats from both domains, provides synchronized navigation across modalities, and allows direct inspection of AI-derived segmentations and annotations alongside the original data. This implementation demonstrates that a modular integration of established tools can yield a practical, reproducible, and extensible solution for multimodal data exploration. The proposed viewer serves as a bridge between radiological and pathological workflows and establishes a foundation for future research, validation, and clinical integration of multimodal diagnostic systems.

\section{Methods}
For improving multimodal image evaluation, two requirements must be fullfilled: the modalities must be mapped onto each other, and the linked dataset must be visualized appropriately. Both branches are tackled with our proposed framework. This section presents the methods that are used to link radiology and pathology images onto each other, as well as the visualization tools to visualize the linked datasets.

\subsection{Data preprocessing}
The integration of both modalities requires preprocessing, including DICOM conversion of Whole Slide Images (WSIs) and definition of their anatomical origin.
\subsubsection{Dicom conversion}
A unified data format is essential for integrating multiple imaging modalities. In radiology, DICOM is the established standard, and with the DICOM-WSI extension it now also supports pathology \cite{dicom_ps3_3_wsi}. Our pipeline therefore requires both modalities in DICOM format and linked through the corresponding DICOM tags \footnote{\url{https://dicom.nema.org/medical/dicom/current/output/html/part18.html}}. To achieve this, we employ the wsi-dicomizer\footnote{\url{https://github.com/imi-bigpicture/wsidicomizer}}, which is also natively integrated into other multimodal imaging platforms such as Kaapana \cite{dicomkaapana}. The tool can be used within Kaapana or independently, supported by the comprehensive documentation.

\subsubsection{Localization of WSI in radiology}
Once both modalities are available in the Picture Archiving and Communication System (PACS), the mapping pipeline can be initiated after verifying data completeness. A key step is aligning WSIs with their corresponding radiological images by resolving the anatomical region. The routine evaluates multiple DICOM tags: \textit{BodyPartExamined}, \textit{Anatomic Region Sequence}, \textit{Admitting Diagnoses Code Sequence} and parses textual fields (\textit{StudyDescription}, \textit{SeriesDescription}, \textit{ProtocolName}, ...) if codes are missing. Free-text entries are normalized to canonical organ labels (heart, prostate, ...) using master JSON derived from controlled DICOM concept lists. The resolver prioritizes exact matches, then broader anatomy groups, and supports substring matching for common label variants.

\subsection{Segmentation pipeline}
After identifying the anatomical site from the WSI file, the corresponding region in the radiological image is determined. The radiological volume is converted to Neuroimaging Informatics Technology Initiative (NIfTI)-format format using dcm2niix \cite{dcm2niix} and, together with the identified body part, passed to TotalSegmentator\cite{totalsegmentator2023} for segmentation . The resulting mask is then converted via itkimage2segimage\footnote{\url{https://qiicr.gitbook.io/dcmqi-guide/opening/cmd_tools/seg/itkimage2segimage}} into a single DICOM-SEG object referencing the original MR images, using dcmqi tooling and a generated meta.json \cite{dcmqi}. The resulting multimodal dataset includes (i) the original radiological DICOM volume, (ii) the DICOM-WSI file, (iii) the anatomical site, and (iv) a dicom segmentation (DICOM SEG) object delineating the organ or biopsy subregion. This step creates a standardized, spatially coherent link between the radiological image, its anatomical context, and the corresponding histopathology. For organs such as kidneys or lungs, a post-processing step merges left and right masks into a single binary mask by voxelwise union, ensuring consistent downstream handling with one segment per anatomical target.

\begin{figure}[h]
  \centering
  \includegraphics[width=\textwidth]{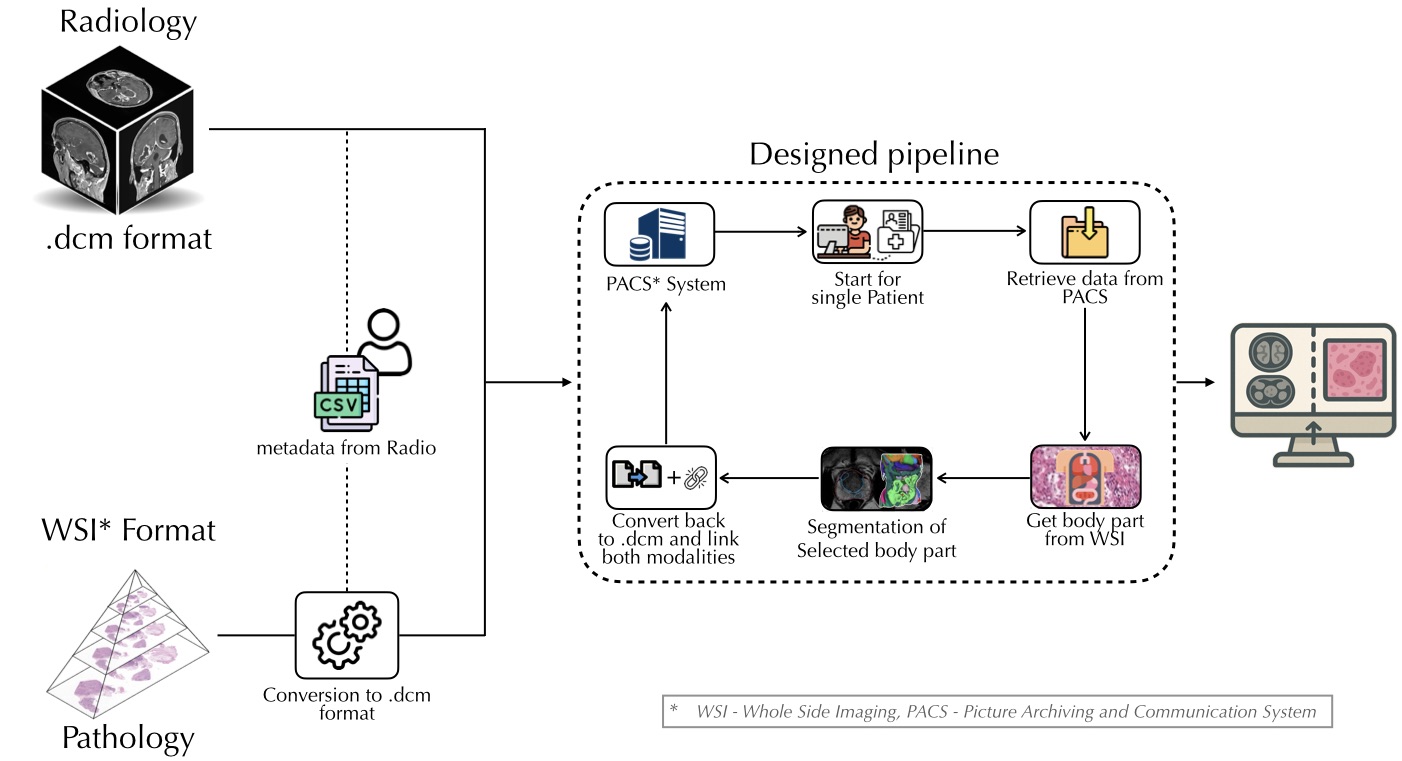}
  \caption{End-to-end pipeline. The flow runs from the user action in the web UI to a DICOM SEG created by dcmqi, with SM-driven body-part inference and master JSON mapping in the middle.}
  \label{fig:reliability}
\end{figure}

\subsection{Combined Viewer}
Navigation across radiological and pathological image viewers typically requires two separate systems, each optimized for its modality. Radiology viewers like Open Health Imaging Foundation (OHIF)-viewer \cite{ohif2020} offer advanced tools for volumetric DICOM studies, while pathology viewers such as SLIde Microscopy (SLIM)-viewer provide interfaces for exploring high-resolution WSIs \cite{slim_idc}. However, cross-modality assessments in separate environments hinder efficient correlation between findings, requiring manual synchronization of spatial context and metadata. To address this, we developed a unified viewer that integrates both modalities within a coordinated web interface. The system embeds the OHIF radiology viewer and SLIM microscopy viewer in a synchronized split-screen layout, enabling interactive, side-by-side exploration of linked regions of interest while preserving full functionality of both viewers. Radiology data such as Magnetic Resonance Imaging (MRI) are stored as DICOM series, and microscopy images (WSI, TIFF, or proprietary formats) are standardized to DICOM via an optional conversion step. All subsequent operations use standard DICOM services and objects. The overall architecture comprises three main components:
\begin{itemize}
    \item \textbf{PACS and services}: A dcm4chee-arc\footnote{\url{https://github.com/dcm4che/dcm4chee-arc-light}} instance can be queried via DICOM web\footnote{\url{https://www.dicomstandard.org/using/dicomweb}} protocols.
    \item \textbf{Backend API}: A java script service with endpoints for (i) study retrieval and packaging, (ii) body-part recognition and concept mapping, and (iii) segmentation pipelines producing DICOM SEG objects \cite{dcmqi}.
    \item \textbf{Web application}: A React-based client lists PACS studies, launches segmentation tasks, and opens the split viewer. OHIF displays radiology data, and SLIM renders microscopy data \cite{ohif2020,slim_idc}. The interface is served by NGINX, which also proxies API and archive traffic.
\end{itemize}

On the client, the study table lists PACS studies. The user selects a radiological study and as default setting, the OHIF viewer opens as entry point to examination for the selected subject.
\subsection{Human Interaction}
Interactive linking between radiological and pathological data is implemented entirely within the web client. The interface captures user selections on radiological images and dynamically loads the corresponding pathology view, enabling synchronized cross-modal navigation. The clickable segmentation overlay is generated directly in the browser by sampling the Cornerstone canvas RGBA buffer after each OHIF rendering event and identifying segmentation pixels using a predefined red-dominant color criterion. A stack-based flood-fill algorithm groups adjacent pixels into contiguous regions, records their bounds, and converts them to screen coordinates based on the current canvas scale. For each region, a semi-transparent, dashed rectangle with a numeric label is rendered as an interactive target without obscuring the underlying image or interfering with OHIF tools \cite{ohif2020}. When selected, the client queries the PACS to locate the corresponding WSI study, retrieves the appropriate images, and displays it in the SLIM viewer on the left while maintaining the OHIF view on the right \cite{slim_idc}. This interaction model enables direct navigation from localized radiological findings to corresponding histopathological data, supporting efficient and spatially consistent multimodal exploration.
\subsection{Implementation}
The Java script backend service calls an PACS endpoint (here we use the dcm4chee-arc PACS) using DICOM web. The anatomic master JSON is mounted into the container and loaded at startup. External tools are invoked with child processes. The service runs inside a container that installs dcm2niix from the system repository \cite{dcm2niix}, \textit{TotalSegmentator} and \textit{SimpleITK} inside a Python virtual environment \cite{totalsegmentator2023}, and dcmqi from a release archive so that itkimage2segimage is available within the pipeline \cite{dcmqi}.
\section{Results}
We verified that MRI studies can be discovered and downloaded through the API, that WSI metadata correctly identifies the intended organ in typical cases, and that the Systematized Nomenclature of Medicine – Clinical Terms (SNOMED CT) resolver returns consistent codes for common targets. The pipeline produces a DICOM SEG for the selected organ and uploads it to the archive. The user interface reflects the outcome and provides clear prompts that indicate what was generated. The archive is always addressed through the internal hostname inside the compose network, which avoids connection errors. The master JSON is mounted with an absolute bind path to prevent inadvertent directory mounts. Python packages are isolated in a virtual environment to honor Debian’s external management policy.

\begin{figure}[H]
  \centering
  \includegraphics[width=\textwidth]{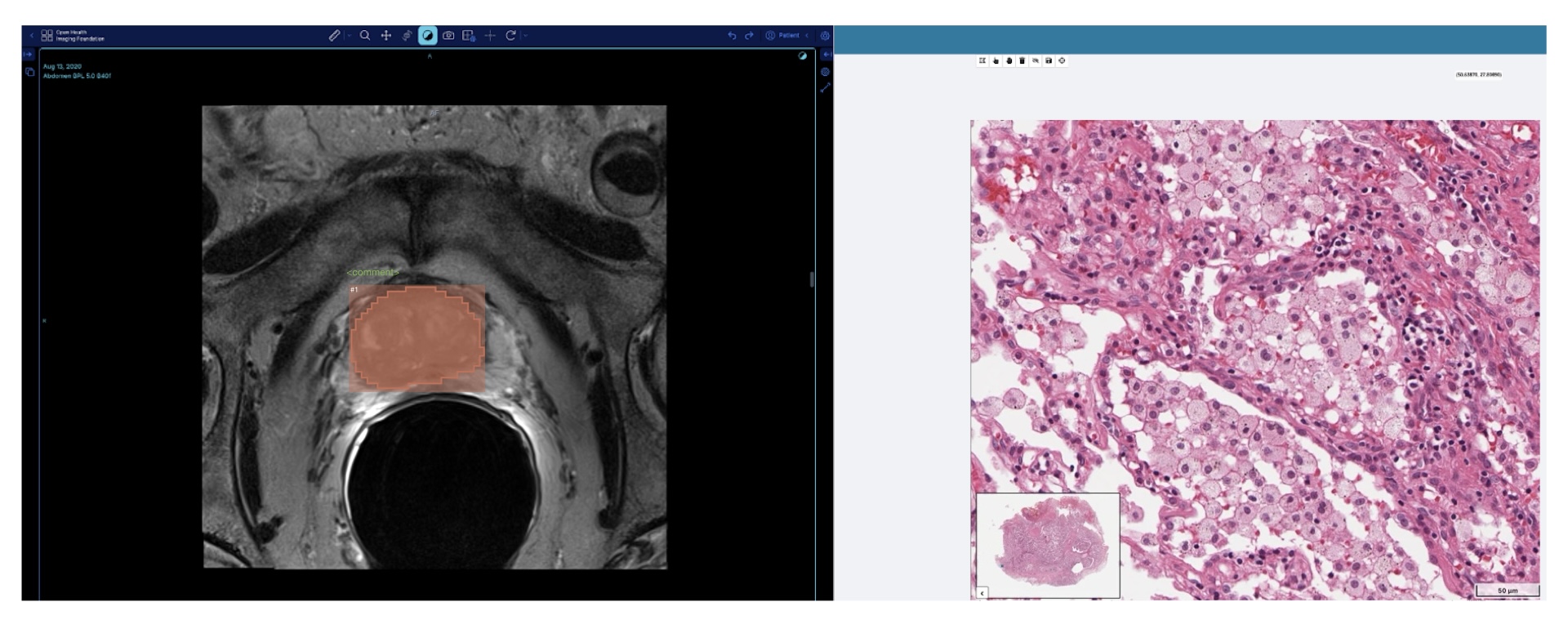}
  \caption{The screenshot of the combined split-viewer which demonstrates simultaneous viewing of images with auto-segmented "Prostate" along with clickable overlay box}
  \label{fig:reliability}
\end{figure}

\section{Discussion}
We developed a modular and extensible pipeline that integrates radiological and pathological imaging within a unified, DICOM-based framework. Using DICOM as the sole input format enables seamless integration with existing PACS infrastructures and clinical data environments. All components are containerized with Docker, ensuring reproducibility and flexible extension of the processing chain. Beyond organ-level mapping, users can incorporate advanced registration methods for fine-grained spatial alignment. The visualization layer follows the same modular design, allowing interchangeable viewer components such as SLIM, OHIF, or other viewers like QuPath\footnote{\url{https://qupath.github.io/}} depending on the use case or institutional preference. This architecture ensures interoperability and adaptability across research and clinical settings. In summary, it provides a robust foundation for collaborative diagnostics and multimodal imaging research.\\
\newpage
\noindent\textbf{\large Acknowledgement}\\

\noindent\textbf{Funding}\\
This work was funded by the German Cancer Research Center CORE-funding program (grant-number: 1010001158/1010001159).\\

\noindent\textbf{Code availability statement}\\
The method will be integrated in the Kaapana toolkit (\url{https://www.kaapana.ai}) and as standalone package \url{https://github.com/MIC-DKFZ/combinedmodalityviewer}.\\

\noindent\textbf{CRediT authorship contribution statement}\\
\textbf{NPR:} Methodology, Software, Investigation, Writing - review \& editing; \textbf{TJB:} Resources, Funding acquisition, Writing - review \& editing; \textbf{PN:} Funding acquisition, Writing - review \& editing; \textbf{MN:} Funding acquisition, Writing - review \& editing; \textbf{KMH:} Resources, Funding acquisition, Writing - Review \& Editing; \textbf{MF:} Conzeptualization, Methodology, Validation, Supervision, Funding acquisition, Writing - original draft; \textbf{CW:} Conzeptualization, Methodology, Supervision, Funding acquisition, Project Administration, Writing - review \& editing.

\printbibliography
\end{document}